\documentclass[twocolumn,pra,superscriptaddress]{revtex4}
\usepackage{graphicx}

\begin{document}

\title {The generator coordinate method in time-dependent density-functional 
theory: memory made simple}

\author {E. Orestes}
\affiliation{Departamento de Qu\'{\i}mica e F\'{\i}sica Molecular, Instituto de Qu\'{\i}mica de S\~ao Carlos,
Universidade de S\~ao Paulo, Caixa Postal 780, S\~ao Carlos, S\~ao Paulo 13560-970, Brazil}
\affiliation {Department of Physics and Astronomy, University of Missouri, Columbia, Missouri 65211, USA}
\affiliation {Departamento de F\'{\i}sica e Inform\'atica, Instituto de F\'{\i}sica de S\~ao Carlos,
Universidade de S\~ao Paulo, Caixa Postal 369, S\~ao Carlos, S\~ao Paulo 13560-970, Brazil}

\author {K. Capelle}
\affiliation {Departamento de F\'{\i}sica e Inform\'atica, Instituto de F\'{\i}sica de S\~ao Carlos,
Universidade de S\~ao Paulo, Caixa Postal 369, S\~ao Carlos, S\~ao Paulo 13560-970, Brazil}

\author {A. B. F. da Silva}
\affiliation {Departamento de Qu\'{\i}mica e F\'{\i}sica Molecular, Instituto de Qu\'{\i}mica de S\~ao Carlos,
Universidade de S\~ao Paulo, Caixa Postal 780, S\~ao Carlos, S\~ao Paulo 13560-970, Brazil}

\author {C. A. Ullrich}
\affiliation {Department of Physics and Astronomy, University of Missouri, Columbia, Missouri 65211, USA}

\date{\today}

\begin{abstract}
The generator coordinate (GC) method is a variational approach to the quantum many-body problem in which
interacting many-body wave functions are constructed as superpositions of (generally nonorthogonal)
eigenstates of auxiliary Hamiltonians containing a deformation parameter.
This paper presents a time-dependent extension of the GC method as a new approach to improve existing approximations of the
exchange-correlation (XC) potential in time-dependent density-functional theory (TDDFT). The time-dependent
GC method is shown to be a conceptually and computationally simple tool to build memory effects into any existing adiabatic
XC potential. As an illustration, the method is applied to driven parametric oscillations of two interacting electrons
in a harmonic potential (Hooke's atom). It is demonstrated that a proper choice of time-dependent generator coordinates in
conjunction with the adiabatic local-density approximation reproduces the exact linear and nonlinear two-electron
dynamics quite accurately, including features associated with double excitations that cannot be captured by TDDFT
in the adiabatic approximation.
\end{abstract}

\maketitle

\section{Introduction} \label{sec:intro}

The development and applications of time-dependent density-functional theory
(TDDFT) \cite{Runge1984,Marques2004} have enjoyed exponential growth in the last few years.
The most successful applications have so far been in the linear or perturbative regime,
to calculate excitation energies and optical spectra of complex molecular systems \cite{furche}.
Other areas of interest for TDDFT are the nonlinear, nonperturbative electron dynamics of
atoms and molecules driven by intense laser fields, and quantum transport through molecular junctions
(see Ref. \cite{TDDFTbook} for a comprehensive review of formalism and applications of TDDFT).

The central quantity in TDDFT, the time-dependent exchange-correlation (XC) potential $v_{\rm xc}(\mathbf{r},t)$,
has to be approximated in practice,
and the success (or the failure) of a TDDFT calculation depends critically on the quality of the chosen XC functional.
The most widely used approximation for the XC potential in TDDFT is the adiabatic
approximation:
\begin{equation}
v_{\rm xc}^{\rm adiabatic}(\mathbf{r},t)=v_{\rm xc}^{\rm static}[n({\bf r},t)] \:,
\end{equation}
where some given XC potential $v_{\rm xc}^{\rm static}[n]$ from static DFT is evaluated at the time-dependent density
$n({\bf r},t)$. This idea seems certainly reasonable for systems that are slowly varying in time.
The most common example of this approach is the adiabatic local-density approximation (ALDA) where, in addition to
a slow change over time one also assumes slow spatial variations.

The adiabatic approximation means that all functional dependence of $v_{\rm xc}({\bf r},t)$ on prior time-dependent
densities $n({\bf r}',t')$, $t'<t$, is ignored. Neglecting the retardation implies frequency-independent
and real XC kernels in linear response \cite{Grosskohn, Petersilka}.
This approach is known to work well for excitation processes in many-body
systems that have a direct counterpart in the Kohn-Sham system, such as atomic and molecular single-particle
excitations. On the other hand, for more
complicated processes such as double or charge-transfer excitations the ALDA can fail dramatically
\cite{Dreuw,Maitra1,Maitra2}. We mention that the well-known failure of the ALDA to reproduce optical absorption
spectra in extended systems \cite{Botti2007} has less to do with the neglect of retardation, but is due to
the wrong spatial long-range behavior of the linear-response XC kernel.

Another well-known class of problems where the adiabatic approximation fails are correlated double photoionization
processes in atoms and molecules (for a review, see \cite{ullrichbandrauk}).
The most notorious case is the direct double ionization of helium \cite{fittinghoff}, which has been a persistently nagging
problem for  TDDFT in any of the standard XC approximations \cite{lappas,Lein2005,mundt}.

Systematic studies of plasmon-like charge-density oscillations in quantum wells and quantum rings
\cite{Wijewardane,DAgosta,UllrichTokatly,Ullrich2006} have uncovered further examples where it is necessary to go beyond the ALDA.
Memory effects have been shown to be crucial in these examples, in particular when
multiple excitations are involved.

A fully-fledged treatment of memory and retardation in TDDFT presents many technical difficulties, although quite a bit of
progress has been made recently. The most fruitful approaches so far are based on ideas of hydrodynamics and elasticity theory,
where the memory resides in the dependence of a fluid element's motion and deformation on its initial position, and where
retardation effects cause the electron liquid to be subject to viscoelastic stresses
\cite{VK,Bunner, VUC,UllrichVignale,Kurzweil,Tokatly}. Applications of these ideas to polymers \cite{Faassen1,Faassen2} and
quantum wells \cite{UllrichVignale2} have met with some success, but there are also
fundamental problems. A particularly disquieting situation arises in finite systems,
where the hydrodynamic TDDFT approach results in excitation energies with unphysical imaginary parts \cite{Ullrich2006,UllrichBurke}.

The purpose of this paper is to present a new, simple and intuitive approach to retardation effects in TDDFT.
The approach is based on an extension the so-called generator coordinate (GC) method, which
was originally introduced in nuclear physics by Wheeler and coworkers \cite{Hill1953,Griffin1957}. The GC method
expresses the many-body wavefunction in terms of a superposition of states arising from
deformed auxiliary Hamiltonians, using
a variational optimization of the expansion coefficients.
In the present paper, we derive and apply a time-dependent generalization of the GC method, combining it with TDDFT.
The result is an approximation to the time-dependent many-body wave function, built from Slater determinants which
come from deformed time-dependent Kohn-Sham Hamiltonians. We will show that by introducing a
class of ``deformations'' involving the previous history of the system, a cheap and simple way to simulate
retardation effects is obtained.

The paper is organized as follows. Section \ref{Framework} first reviews the basic GC formalism,
and then introduces the time-dependent GC method in the context of TDDFT.
Our model system, Hooke's atom, is briefly discussed in Section \ref{sec:Hooke}.
In Section \ref{Res+Dis} we present applications of the GC-TDDFT approach to
a simple model system (parametric oscillations of two electrons in a harmonic potential, also known as Hooke's atom).
Section \ref{Sum} gives a summary and conclusions. Some technical details are given in the appendices.
We use atomic (Hartree) units everywhere, with $e=m=\hbar=1$.

\section{Formal Framework}
\label{Framework}

\subsection{The GC method}
\label{GCM}

In the GC method, approximations for the full many-body wave function $\Psi({\bf x}_1,...,{\bf x}_N)$ of
a system of $N$ electrons, where ${\bf x}=({\bf r},s)$ denotes spatial coordinate and spin,
are constructed as follows \cite{Griffin1957}:
\begin{equation} \label{staticansatz}
\Psi({\bf x}_{1},...,{\bf x}_{N})=\int d\alpha f(\alpha)\Phi(\alpha,{\bf x}_1,...,{\bf x}_N)\label{GCwf}.
\end{equation}
Here, $\Phi(\alpha,{\bf x}_1,...,{\bf x}_N)$ represents a set of (usually noninteracting)
seed functions which depend on a deformation parameter $\alpha$, the generator coordinate. This dependence
can be explicit (if $\alpha$ is a parameter in an analytic expression for $\Phi$) or implicit (e.g., if the $\Phi$ are
eigenstates of a Hamiltonian which contains the parameter $\alpha$).

The weight function $f(\alpha)$ is determined variationally.
Variation with respect to $f^*(\alpha)$ of the normalized many-body expectation value
$\langle \Psi | \hat{H} | \Psi \rangle/\langle \Psi| \Psi \rangle$,
where $\hat{H}$ is the Hamiltonian of the fully interacting system, leads to the so-called Griffin-Hill-Wheeler (GHW)
equation:
\begin{equation}\label{GHWeq}
\int d\alpha'\left[ K(\alpha,\alpha') - ES(\alpha,\alpha') \right] f(\alpha')=0,
\end{equation}
where $K(\alpha,\alpha')=\langle\Phi(\alpha)|\hat{H}|\Phi(\alpha')\rangle$ and
$S(\alpha,\alpha')=\langle\Phi(\alpha)|\Phi(\alpha')\rangle$ are the Hamiltonian and overlap matrix elements of the
given seed functions. Solution of the GHW equation (\ref{GHWeq}) yields a set of
energy eigenvalues $E$ and associated eigenfunctions $f(\alpha)$ which are used in Eq. (\ref{GCwf}) to determine
the corresponding many-body wave functions and all observables following therefrom.

In the original work of Wheeler {\em et al}. \cite{Hill1953,Griffin1957}, $\Psi$ was a nuclear many-body wave
function, and the generator coordinate was a parameter describing the deformed shape of the potential due
to the collective behavior of the nucleons. The seed functions were obtained from a simplified Schr\"odinger-like
equation featuring a deformed nuclear potential.

Over the years, the GC method has found widespread use in a large variety of
problems in nuclear and electronic many-body physics \cite{Wong1975,Chattopadhyay1978,Johansson1978}.
In quantum chemistry, the GHW variational equations were used to develop
high-quality Hartree-Fock and Dirac-Fock single-particle basis functions
\cite{Mohallen1986,daSilva1989,Jorge96a,Jorge96b}.
Expressions similar to the GC {\em ansatz} (\ref{staticansatz})
have recently also been used by Alon, Streltsov and Cederbaum \cite{Alon2005}
in order to build symmetries into the many-body wave function $\Psi$ that
are not present in the seed wave functions $\Phi$; and by Pan, Sahni and Massa
\cite{Pan2004} with the aim to allow more general variations of the wave
function during variational calculations.

\subsection{Static GC-DFT}
\label{GC-DFT}

Recently, a connection between the GC method and static DFT was established by Capelle \cite{Capelle2003}.
The idea is to express the electronic many-body wave function (\ref{GCwf}) as a weighted superposition
of Kohn-Sham (KS) Slater determinants $\Phi_{\rm KS}$, constructed with KS orbitals $\varphi_j^\alpha$:
\begin{equation} \label{TDansatz}
\Psi({\bf x}_1,...,{\bf x}_N)=\int d\alpha f(\alpha)\Phi_{\rm KS}
(\alpha,{\bf x}_1,..., {\bf x}_N)\label{GCDFTwf}.
\end{equation}
The generator coordinate is introduced, like in the original proposal of Wheeler \textit{et al.}
\cite{Hill1953,Griffin1957}, as a deformation parameter of the potential, in this case of the XC potential.
The KS orbitals $\varphi_j^\alpha$, $j=1,\ldots,N$, following from
\begin{equation}\label{staticKS}
\left[-\frac{\nabla^2}{2} + v_{\rm ext}({\bf r}) + v_{\rm H}({\bf r}) + v_{\rm xc}^{\alpha}({\bf r}) \right]
\varphi_j^{\alpha}({\bf r}) = \epsilon_j^\alpha \varphi_j^{\alpha}({\bf r}) \:,
\end{equation}
where $v_{\rm ext}$ and $v_{\rm H}$ are the external and Hartree potentials,
used to construct the Hamiltonian and overlap matrix elements $K(\alpha,\alpha')$ and $S(\alpha,\alpha')$
in the GHW equation (\ref{GHWeq}). In practice, the GC-DFT wave function (\ref{GCDFTwf}) is constructed using a finite
set of deformation parameters $\alpha$ (usually less than 10), forming a mesh of points which is used to
discretize the GHW integral equation (\ref{GHWeq}).
Here and in the following, we only consider systems that are nonmagnetic everywhere and thus ignore the spin index.

A first illustrative numerical calculation of this new approach, showing its viability and potential,
was made in Ref. \cite{Capelle2003} for the Helium isoelectronic series. In this case, a single set of deformation
parameters was tested together with a simple
functional, the X-only LDA, to obtain ground-state energies and the weight functions.
Through Eq. (\ref{GCDFTwf}) an approximation to the many-body wave function was established and the expectation values
of the operator $r^n$, with $n=-2,-1,0,1,2$, were calculated.

The results of Ref. \cite{Capelle2003}
show that the ground-state energies obtained are considerably better than those calculated  from both the LDA
generator functionals themselves and more sophisticated gradient-corrected XC functionals. The many-body wave functions were obtained
with almost no additional numerical effort, yielding expectation values that are close to those given by other methods.
Thus, the GC method is a cheap and conceptually simple way to improve existing XC functionals. The method is, of course, not restricted
to the helium isoelectronic series. Moreover, it can also be used to extract excitation energies. Work along these lines will
be reported elsewhere \cite{Orestes}.

\subsection{Time-dependent GC-DFT}
\label{GC-TDDFT}
Let us now extend the GC-DFT approach into the time domain. Our starting point is a generalization of
Eq. (\ref{GCwf}): we write the time-dependent many-body wave function $\Psi$ as a linear superposition of
time-dependent KS (TDKS) Slater determinants $\Phi_{\rm KS}$ depending on the generator coordinate $\alpha$:
\begin{equation}\label{GCTDDFTwf}
\Psi({\bf x}_1,...,{\bf x}_N,t)=\int d\alpha f(\alpha,t)
\Phi_{\rm  KS}(\alpha,{\bf x}_1,..., {\bf x}_N,t) .
\end{equation}
The weight function $f(\alpha,t)$ is taken to be time-dependent and complex. We consider situations
where the system is in a stationary state for $t\le t_0$. The time evolution of $\Psi$ for $t>t_0$
thus represents an initial value problem, where $\Psi$, $\Phi_{\rm KS}$ and $f$  at the initial time $t_0$
are obtained from the static GC-DFT method described above.

Like in the static case, the KS Slater determinants are built from single-particle orbitals
$\phi^\alpha_j({\bf r},t)$, $j=1,\ldots,N$, where
\begin{equation} \label{TDKS}
\left[ -\frac{\nabla^2}{2} + v_{\rm ext}({\bf r},t) + v_{\rm H}({\bf r},t) + v_{\rm xc}^\alpha({\bf r},t)-
i\frac{\partial}{\partial t}  \right] \phi_{j}^{\alpha}({\bf r},t) =0.
\end{equation}
The initial conditions are $\phi_{j}^{\alpha}({\bf r},t_0)=\varphi_j^\alpha({\bf r})$ for each $\alpha$,
see equation (\ref{staticKS}).

The time-dependent weight functions $f(\alpha,t)$ are determined through a stationary-action principle:
\begin{equation}\label{StationaryAction}
\frac{\delta}{\delta f^*(\alpha,t)}\left[ \int^{t_1}_{t_0} dt'\left\langle \Psi(t')\left|i\frac{\partial}{\partial t'} -
\hat{H}(t') \right| \Psi(t') \right\rangle \right] = 0,\label{VarF's}
\end{equation}
where $t_1$ is an arbitrary upper limit of the time propagation interval. $\hat{H}(t)$ is the time-dependent many-body
Hamiltonian featuring the external one-body potential $v_{\rm ext}({\bf r},t)$.

The solution of Eq. (\ref{VarF's}) leads to the time-dependent GHW (TDGHW) equation:
\begin{equation} \label{TDGHW}
\int d\alpha'\left[ A(\alpha,\alpha',t) + S(\alpha,\alpha',t)i\frac{\partial}{\partial t} \right]
f(\alpha',t)=0,
\end{equation}
where
\begin{equation} \label{actionmatrix}
A(\alpha,\alpha',t)=\left\langle \Phi^*_{\rm KS}(\alpha,t) \left|i\frac{\partial}{\partial t}-\hat{H}(t)\right|
\Phi_{\rm KS}(\alpha',t)\right\rangle
\end{equation}
are the KS action matrix elements, and
\begin{equation}\label{overlapmatrix}
S(\alpha,\alpha',t)=\left\langle \Phi^*_{\rm KS}(\alpha,t)| \Phi_{\rm KS}(\alpha',t)\right\rangle
\end{equation}
are the time-dependent KS overlap matrix elements. In Appendix \ref{AppendixA} we show that the TDGHW equation
satisfies two key properties: it reduces to the static GHW equation (\ref{GHWeq}) in the limit where $v_{\rm ext}$ is
time-independent, and it conserves the norm of the total wavefunction (\ref{GCTDDFTwf}).

In practice, one first chooses a mesh of $\alpha$-values and
obtains the initial KS orbitals $\phi_{j}^\alpha({\bf r},t_0)$ and the weight functions $f(\alpha,t_0)$
from a static GC-DFT calculation. One then propagates the TDKS equation (\ref{TDKS}) for each KS orbital
and each value of $\alpha$, from $t_0$ to $t_1$. Then, the
matrix elements (\ref{actionmatrix}) and (\ref{overlapmatrix}) are formed, and the TDGHW equation can be solved by
time propagation, which yields the time-dependent weight functions $f(\alpha,t)$. The final step is to calculate
the observables of the system under study.

In Appendix \ref{AppendixB} we discuss our numerical approach to solve the discretized TDGHW equation using
the Crank-Nicholson algorithm, which guarantees a unitary time evolution of $\Psi(t)$.

\section{Hooke's atom: Model and Observables}
\label{sec:Hooke}

To illustrate the GC-TDDFT approach, we consider a system of two interacting electrons confined
by a spherically symmetric harmonic potential $kr^2/2$, known as Hooke's atom.
The great advantage of this system is that the associated two-body Schr\"odinger equation separates
into center-of-mass and relative coordinates, and can therefore be solved in a straightforward manner, even in the
time-dependent case
\cite{Keisner1962,Laufer1986,Kais1993,Taut1994,Filippi1994,DAmico1999,Hessler1999,Hessler2002}. This
allows us to compare our approximate TDDFT and GC-TDDFT results with exact numerical solutions of the full Schr\"odinger equation,
at a relatively small computational cost.

In the following, we shall consider Hooke's atom driven by a time-dependent spring constant
\begin{equation} \label{TDspring}
k(t)=[1+A s(t) \sin(\omega t)]k_0 \:,
\end{equation}
where $k_0$ is the spring constant of the initial ground state,
$A$ and $\omega$ are the amplitude and frequency of the oscillations, and $s(t)$ defines a ``pulse shape''.

Hooke's atom resembles the helium atom in many respects, but there is one important difference:
due to the harmonic confining potential, Hooke's atom has no continuum of unbound states and hence cannot be ionized.
Therefore it cannot be used to address the issue of direct double ionization
\cite{fittinghoff,lappas,Lein2005,mundt}. Nevertheless, Hooke's atom is a useful model system for studying correlated
electron dynamics, including double excitations.

It is known that TDDFT provides the time-dependent density $n({\bf r},t)$ in principle exactly,
but runs into difficulties whenever one is interested in observables that are not easily expressed as explicit
functionals of the density. Thus, a quantity such as the radial expectation value
\begin{equation}
\langle r(t)\rangle=\int d^3r \: n(\mathbf{r},t) \label{ExpR}
\end{equation}
can be easily calculated, since it involves only the total density. By contrast consider the following quantities
for a two-electron system \cite{petersilka1999,ullrich2000}:
\begin{eqnarray}
P^{(0)}(t) & = & \frac{1}{2}\int_{r_{1}<R} d^3r_1 \int_{r_{2}<R} d^3r_2
|\Psi(\mathbf{r}_1,\mathbf{r}_2,t)|^2  \label{p0}\\
P^{(2)}(t) & = & \frac{1}{2}\int_{r_{1}>R} d^3r_1 \int_{r_{2}>R} d^3r_2
|\Psi(\mathbf{r}_1,\mathbf{r}_2,t)|^2  \label{p2}\\
P^{(1)}(t) & = &  1 - P^{(0)}(t) - P^{(2)}(t),
\label{p1}
\end{eqnarray}
where $P^{(0)}$ and $P^{(2)}$ are the probabilities of finding both electrons inside or
outside of an imaginary spherical box of radius $R$ surrounding the system, and $P^{(1)}$ is the probability
of finding one electron inside and the other one outside the box. For a two-electron real atom, these quantities are
ionization probabilities, provided $R$ is large enough.

By the Runge-Gross theorem \cite{Runge1984}, $P^{(0)}$, $P^{(1)}$ and $P^{(2)}$ have the formal property of being
functionals of the time-dependent density. The explicit form of these functionals, however, is unknown, and it is
therefore very difficult to extract $P^{(0)}$, $P^{(1)}$ and $P^{(2)}$ only from $n({\bf r},t)$.
It is common practice (although not formally justified) to replace the full interacting wave function $\Psi$ in
Eqs. (\ref{p0})--(\ref{p1}) by the KS wave function $\Phi_{\rm KS}$ \cite{ullrich2000}. However, it comes as no surprise
that the so-defined KS probabilities fail to capture the essence of the correlated double ionization processes in
helium \cite{lappas}. A method that is wavefunction based, such as GC-DFT, has a much better chance for success.

\section{Results and Discussion}
\label{Res+Dis}

We now present results for parametrically driven oscillations of Hooke's atom, comparing exact solutions of
the time-dependent two-particle Schr\"odinger equations with various levels of TDDFT and time-dependent GC-DFT.

In the following, we use time-dependent spring constants $k(t)$ with a trapezoidal shape for $s(t)$, with a switch-on by a
linear two-cycle ramp, followed by 5 cycles at constant amplitude $A$, and a two-cycle linear switch-off.
Driven by the time-dependent spring constant, the system carries out breathing-mode oscillations
which preserve the initial spherical symmetry. We shall consider driving frequencies $\omega$ in the
vicinity of the lowest resonance at $\omega_0=1.95$ a.u., for $k_0=1$ a.u.

\subsection{ALDA}

\begin{figure}
\includegraphics[angle=-90,width=8cm]{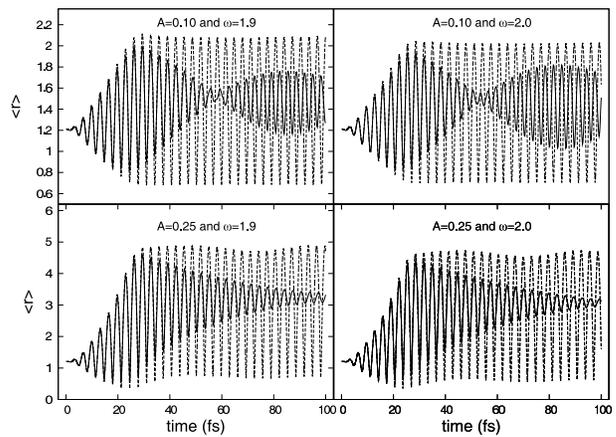}
\caption{$\langle r(t) \rangle$ for a parametrically driven Hooke's atom,
with spring constant $k(t)$ given by Eq. (\ref{TDspring}). Top and bottom panels: $A=0.1$ and 0.25; left and right
panels: $\omega = 1.9$ and 2.0 (above and below resonance). The ``pulse shape'' of $k(t)$ is trapezoidal, with a duration
of 9 cycles. Solid lines: exact calculation. Dashed lines: ALDA.}
\label{ExALDAr}
\end{figure}

\begin{figure}
\includegraphics[angle=-90,width=8cm]{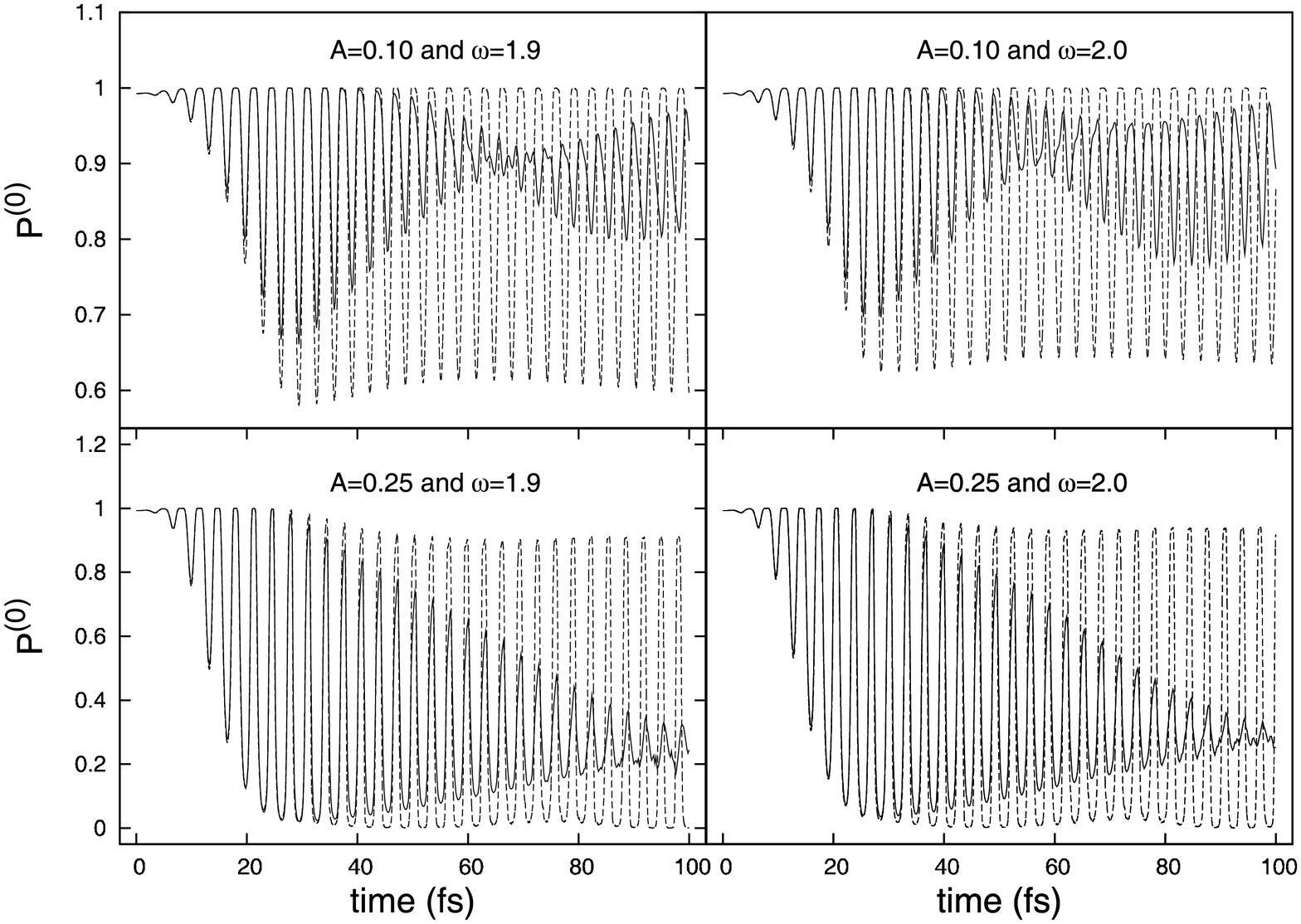}
\caption{
Probability $P^{(0)}$ [Eq. (\ref{p0})] of finding both electrons inside a spherical box of radius $R=2.7$ a.u.,
for the system of Fig. \ref{ExALDAr}. Solid lines: exact calculation. Dashed lines: ALDA.
}
\label{ExALDAp0}
\end{figure}

\begin{figure}
\includegraphics[angle=-90,width=8cm]{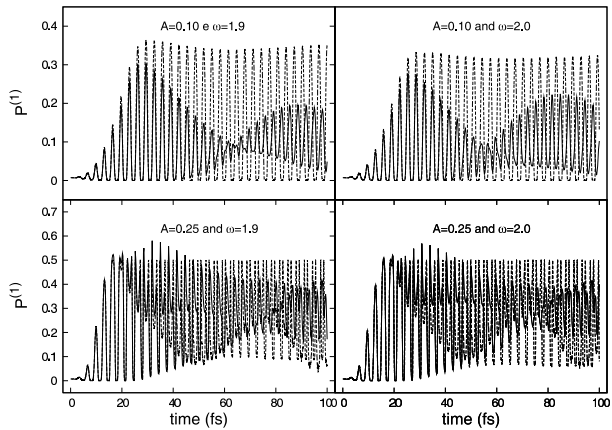}
\caption{
Probability $P^{(1)}$ [Eq. (\ref{p1})] of finding one electrons inside a spherical box of radius $R=2.7$ a.u., and the other one outside,
for the system of Fig. \ref{ExALDAr}. Solid lines: exact calculation. Dashed lines: ALDA.
}
\label{ExALDAp1}
\end{figure}

\begin{figure}
\includegraphics[angle=-90,width=8cm]{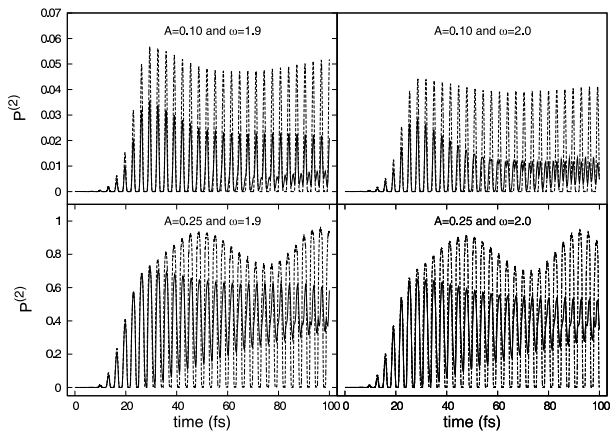}
\caption{
Probability $P^{(2)}$ [Eq. (\ref{p2})] of finding both electrons outside a spherical box of radius $R=2.7$ a.u.,
for the system of Fig. \ref{ExALDAr}. Solid lines: exact calculation. Dashed lines: ALDA.
}
\label{ExALDAp2}
\end{figure}

We first compare (x)-only ALDA with exact results for the time-dependent Hooke's atom.
Fig. \ref{ExALDAr} shows $\langle r(t) \rangle$, and Figs. \ref{ExALDAp0}--\ref{ExALDAp2} present
$P^{(0)}$, $P^{(1)}$ and $P^{(2)}$, respectively, for four different combinations of amplitudes and frequencies
of the driving force constant ($A=0.1,0.25$, and $\omega=1.9, 2.0$). The spherical box used to compute the probabilities $P^{(i)}$
has radius $R=2.7$ a.u.

The ALDA completely fails to reproduce the behavior of the driven Hooke's atom. This can be seen
very clearly from Fig. \ref{ExALDAr}: in ALDA, the radial expectation value oscillates with a constant amplitude after
the trapezoidal pulse is over. By contrast, the exact $\langle r(t) \rangle$ exhibits a beating pattern, where the
amplitude of the oscillations is strongly modulated. Likewise, the probabilities $P^{(0)}$, $P^{(1)}$ and $P^{(2)}$ show
a drastic amplitude modulation, which is absent in ALDA.

The physical origin of the beating patterns in the charge-density oscillations is a
superposition of the dynamics associated with single and multiple excitations. For Hooke's atom,
the beating pattern is particularly pronounced since, due to the harmonic confining potential, the electronic
level separations are nearly equidistant, which makes the coupling to double and higher excitations very efficient.
A similar though less pronounced modulation effect of charge-density oscillations was recently analyzed for two
electrons confined on a two-dimensional quantum strip \cite{Ullrich2006}.

It is a well-known fact that the ALDA fails to capture any type of excitations which does not have a counterpart
in the Kohn-Sham single-particle spectrum, such as double excitations \cite{Maitra1,Maitra2}.
Nevertheless, the sheer magnitude of the error of ALDA
exhibited in Figs. \ref{ExALDAr}--\ref{ExALDAp2} comes somewhat as a surprise.

\subsection{Adiabatic GC-TDDFT}

\begin{figure}
\includegraphics[angle=-90,width=8cm]{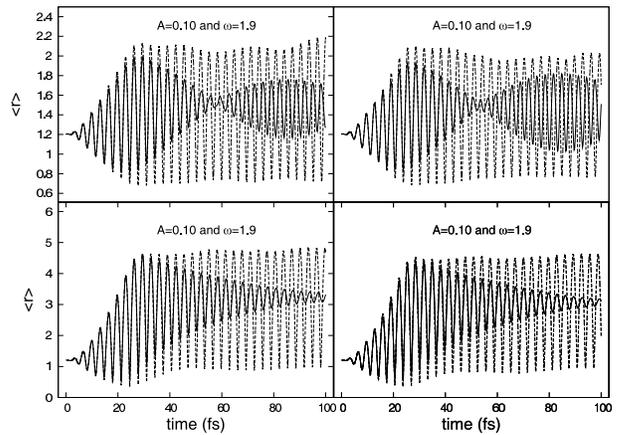}
\caption{
Same as Fig. \ref{ExALDAr}. Solid lines: exact calculation. Dashed lines: adiabatic GC-TDDFT.
}
\label{ExHXCr}
\end{figure}

\begin{figure}
\includegraphics[angle=-90,width=8cm]{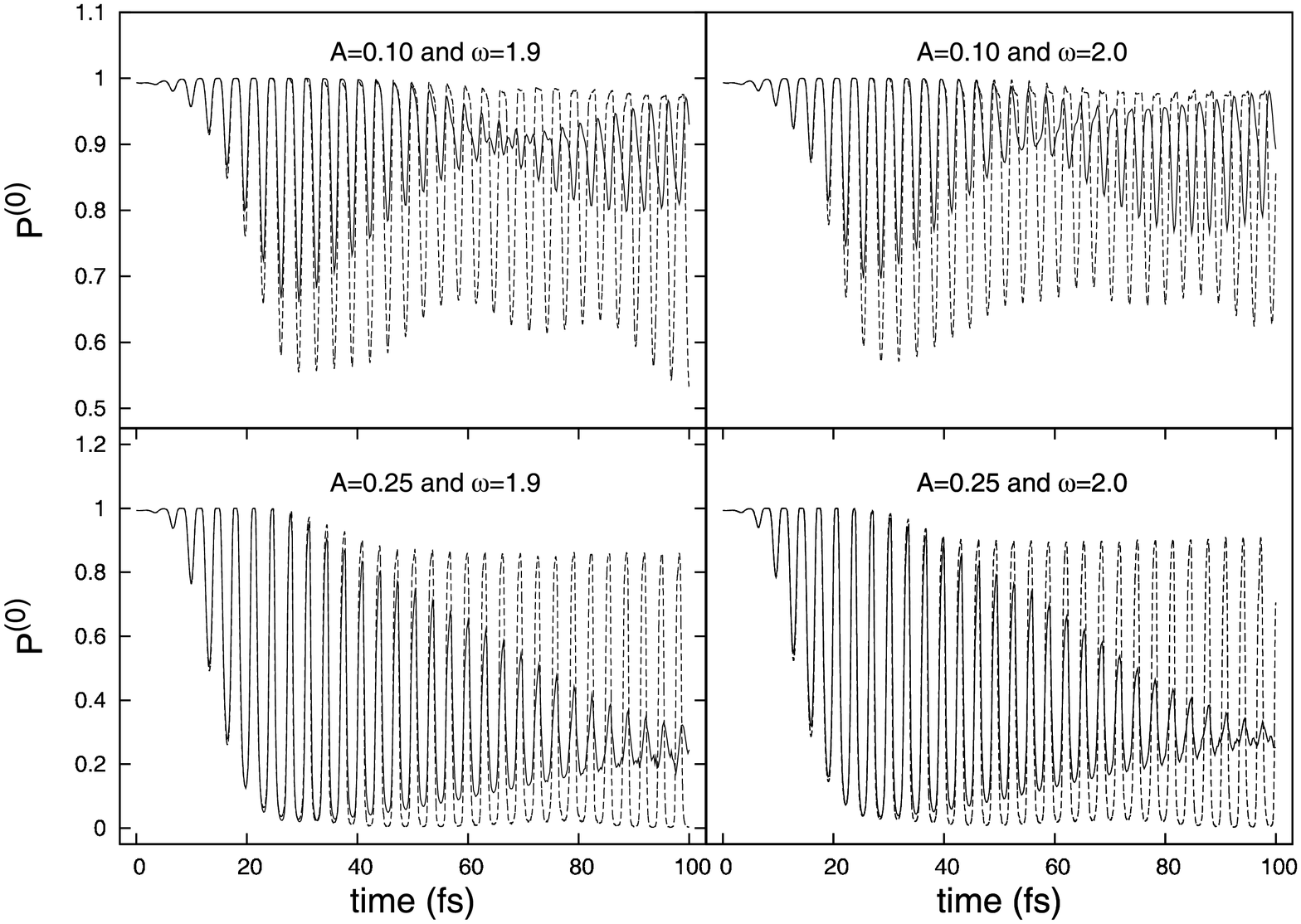}
\caption{
Same as Fig. \ref{ExALDAp0}. Solid lines: exact calculation. Dashed lines: adiabatic GC-TDDFT.
}
\label{ExHXCp0}
\end{figure}

\begin{figure}
\includegraphics[angle=-90,width=8cm]{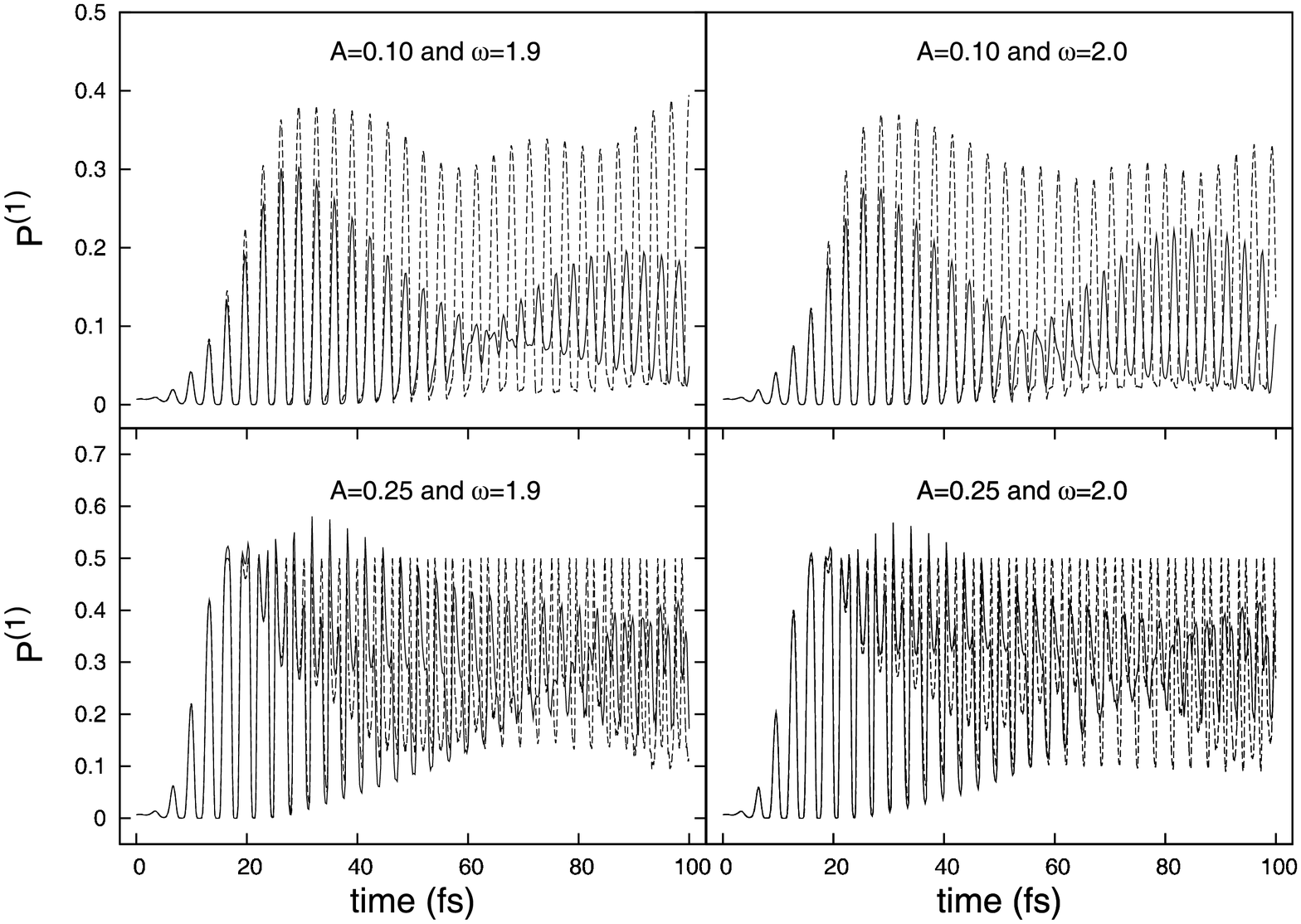}
\caption{
Same as Fig. \ref{ExALDAp1}. Solid lines: exact calculation. Dashed lines: adiabatic GC-TDDFT.
}
\label{ExHXCp1}
\end{figure}

\begin{figure}
\includegraphics[angle=-90,width=8cm]{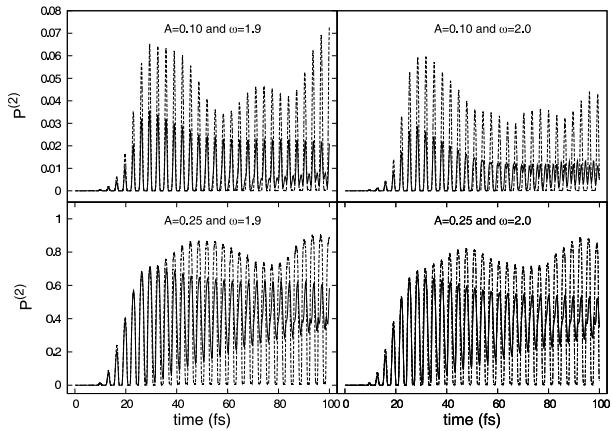}
\caption{
Same as Fig. \ref{ExALDAp2}. Solid lines: exact calculation. Dashed lines: adiabatic GC-TDDFT.
}
\label{ExHXCp2}
\end{figure}

In a first attempt to improve the ALDA description with the GC-TDDFT approach, we consider the following deformed xc potential:
\begin{equation}
v_{\rm xc}^\alpha(\mathbf{r},t)=(\alpha-1) v_{\rm H}({\bf r},t) + \alpha v_{\rm xc}^{\rm LDA}[n({\bf r},t)] \:.
\label{AdVxc}
\end{equation}
For $\alpha=1$, this reduces to the ALDA, and for $\alpha=0$ the xc potential vanishes and the first term cancels the Hartree
potential in the TDKS equation, leaving only the bare confining potential. This describes the situation where one
electron is far away from the center, and the other electron only sees the bare potential. Due to the self-interaction
error, the ALDA does not capture such a scenario correctly, whereas the deformed XC potential
(\ref{AdVxc}), for $\alpha=0$, cancels this error.

We have chosen the set of deformation parameters $\alpha = \{0,1,2,3,4\}$ and the x-only ALDA in Eq. (\ref{AdVxc}).
The results of this adiabatic GC-TDDFT approach for the charge-density oscillations in Hooke's atom are shown in
Figs. \ref{ExHXCr}--\ref{ExHXCp2}, for the same parameters as in Figs. \ref{ExALDAr}--\ref{ExALDAp2}.
We find that the improvement of the adiabatic GC-TDDFT over the pure ALDA is marginal. There is a hint
of an amplitude modulation in all cases, but it is by far insufficient compared to the exact data.

A possible reason for the failure of this method is the peculiar nature of Hooke's atom which prevents ionization.
It seems plausible that in a real atom, where one or both electrons can ionize, the approximate treatment of the
self-interaction error implied in Eq. (\ref{AdVxc}) might be more beneficial.

\subsection{Nonadiabatic GC-TDDFT: memory}

\begin{figure}
\includegraphics[angle=-90,width=8cm]{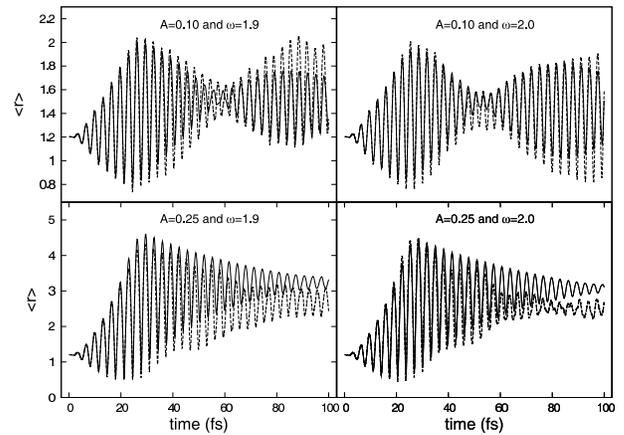}
\caption{
Same as Fig. \ref{ExALDAr}. Solid lines: exact calculation. Dashed lines: nonadiabatic GC-TDDFT.
}
\label{ExHXCdr}
\end{figure}

\begin{figure}
\includegraphics[angle=-90,width=8cm]{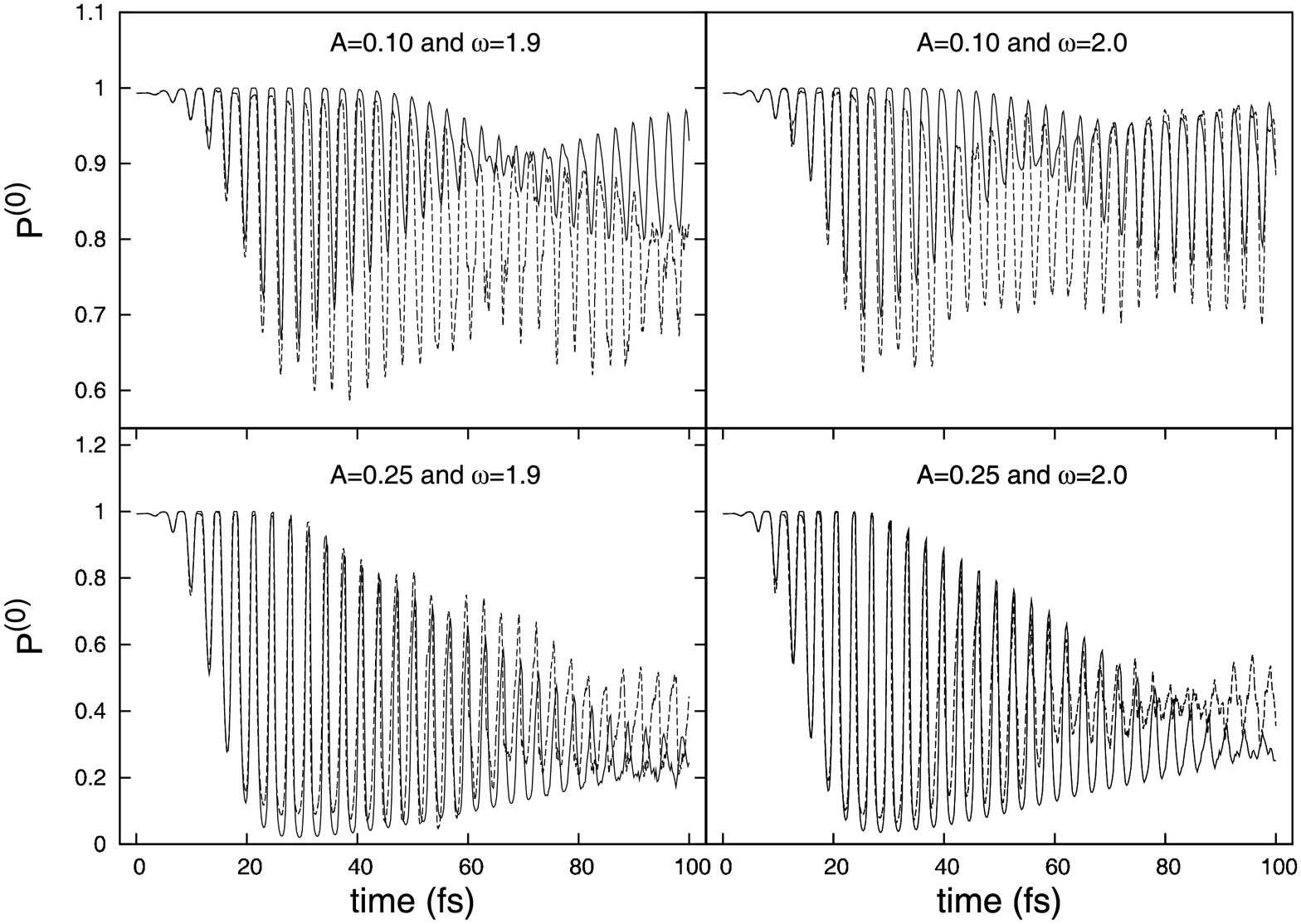}
\caption{
Same as Fig. \ref{ExALDAp0}. Solid lines: exact calculation. Dashed lines: nonadiabatic GC-TDDFT.
}
\label{ExHXCdp0}
\end{figure}

\begin{figure}
\includegraphics[angle=-90,width=8cm]{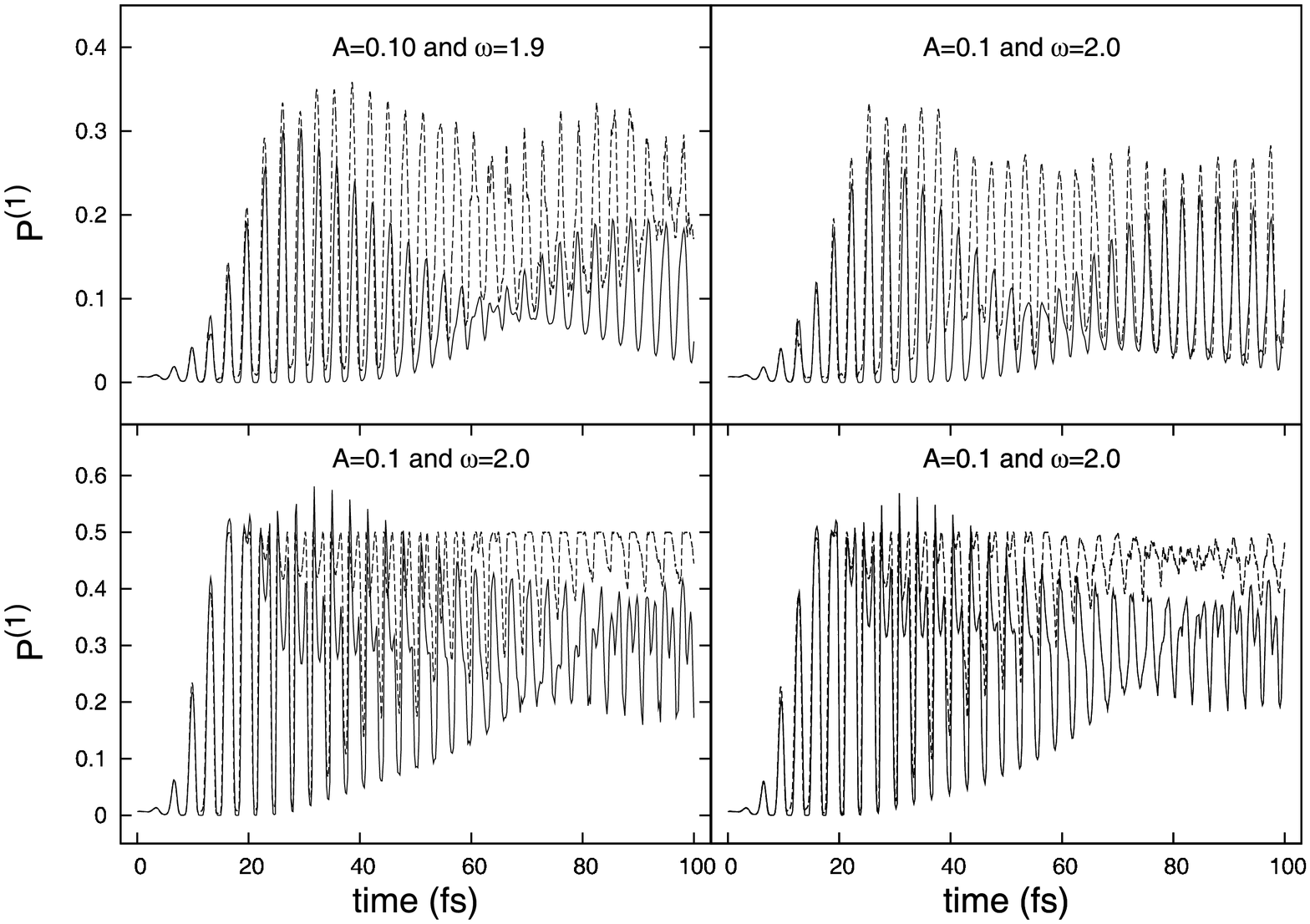}
\caption{
Same as Fig. \ref{ExALDAp1}. Solid lines: exact calculation. Dashed lines: nonadiabatic GC-TDDFT.
}
\label{ExHXCdp1}
\end{figure}

\begin{figure}
\includegraphics[angle=-90,width=8cm]{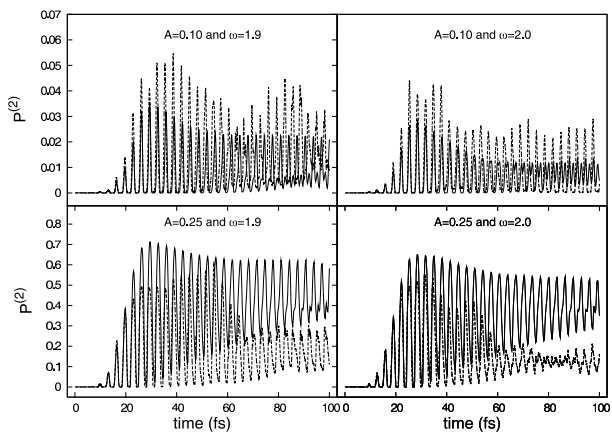}
\caption{
Same as Fig. \ref{ExALDAp2}. Solid lines: exact calculation. Dashed lines: nonadiabatic GC-TDDFT.
}
\label{ExHXCdp2}
\end{figure}

As discussed in the introduction, the adiabatic approximation omits memory and retardation effects, which have
been shown to be crucial for the treatment of double and multiple excitations \cite{Maitra1}. Therefore, we
will now choose a deformation in the GC-TDDFT approach which introduces retardation in a very simple manner. Let
\begin{equation}
v_{\rm xc}^\alpha(\mathbf{r},t)=(\alpha-1) v_{\rm H}({\bf r},t) + \alpha v_{\rm xc}^{\rm LDA}[n({\bf r},t_r)] \:,
\label{NoAdVxc}
\end{equation}
where the retarded time, $t_r$, is given by
\begin{equation} \label{ttilde}
t_r = t - \frac{\alpha T}{8} \:.
\end{equation}
Here, $T=2\pi/\omega$ denotes one cycle of the time-dependent spring constant used to trigger the charge-density oscillations.
Equations (\ref{NoAdVxc}) and (\ref{ttilde}) imply that the LDA xc functional at time $t$ is evaluated using the time-dependent
density at a previous time $t_r$, where $\alpha$ determines how far we go back into the past. Choosing again the set
of deformation parameters $\alpha = \{0,1,2,3,4\}$, we go back at most half a cycle, which corresponds to a phase lag
of $180^{\rm o}$ of the deformed xc potential with respect to the charge-density oscillations.

Results are shown in Figs. \ref{ExHXCdr}--\ref{ExHXCdp2} for the same parameters as in
Figs. \ref{ExHXCr}--\ref{ExHXCp2} and \ref{ExALDAr}--\ref{ExALDAp2}. We obtain an excellent agreement between the
exact and the nonadiabatic GC-TDDFT results for $\langle r(t)\rangle$, where the strong amplitude modulations
of the charge-density oscillations are very well reproduced for all four parameter combinations in Fig. \ref{ExHXCdr}.
The improvement for the probabilities $P^{(i)}$, Figs. \ref{ExHXCdp0}--\ref{ExHXCdp2}, is also very substantial, especially
for the larger amplitude $A=0.25$.

\subsection{Weight Functions}

\begin{figure}
\includegraphics[angle=-90,width=8cm]{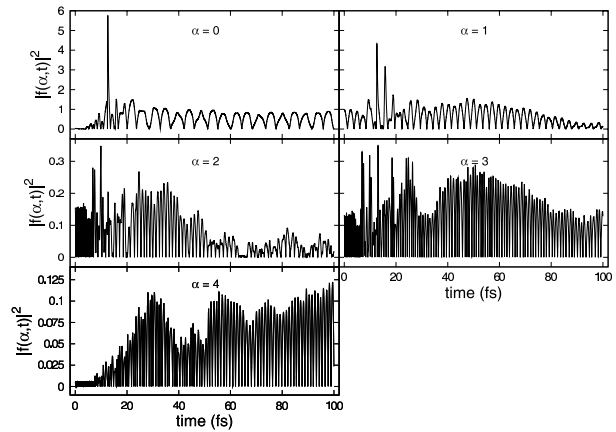}
\caption{Weight functions $|f(\alpha,t)|^2$ for the nonadiabatic GC-TDDFT calculation of the driven Hooke's atom, with
$A=0.1$ and $\omega=1.9$.}
\label{ExHXCdWF}
\end{figure}

In order to analyze the contribution of each TDKS determinant in the approximation of the time-dependent many-body
wave function, $\Psi(\mathbf{r}_{1},...,\mathbf{r}_{N},t)$, we plot the square modulus of the weight
functions $|f(\alpha,t)|^2$ in Fig. \ref{ExHXCdWF}, for the nonadiabatic GC-TDDFT calculation with $A=0.1$ and $\omega=1.9$.
We will focus in particular on the two cases $\alpha = 2$ and 4, which corresponds to a $90^{\rm o}$
and $180^{\rm o}$ phase lag of the xc potential with respect to the driving force.

We observe that $|f(2,t)|^2$ strongly decreases after $t\approx50$ fs, after the end of the driving trapezoidal pulse.
The opposite happens for $|f(4,t)|^2$, which starts out small and picks up after the driving pulse is over.
Taken together, this is exactly what is needed to to produce the observed amplitude modulations of the charge-density
oscillations, with a node approximately at $t\approx50$ fs, see Fig. \ref{ExHXCdr}. The amplitude suppression
happens due to the contribution of the $\alpha=2$ TDKS determinant, where the $90^{\rm o}$ phase lag simulates
dissipation. The subsequent increase of the amplitude requires the contribution of the $\alpha=4$ determinant, where
the $180^{\rm o}$ phase lag simulates an elastic behavior.

A similar way of explaining the amplitude modulations was discussed in Ref. \cite{Ullrich2006}. The time-dependent xc
potential alternatingly acts as a driving and a dissipative force, and is thus able to increase and reduce the amplitude
of the charge-density oscillations.

\section{Summary and Conclusions}
\label{Sum}

The adiabatic approximation, in particular the ALDA, continues to be the workhorse for mainstream
applications in TDDFT, such as molecular excitation spectra. However, as shown by
Maitra {\em et al.} \cite{Maitra1,Maitra2}, the description of more complicated processes
such as multiple or charge-transfer excitations mandates a time-dependent XC functional which includes
memory. Since these excitation processes are extremely relevant in (bio)chemistry,
there is an urgent need to address the question of how to go beyond the adiabatic approximation in TDDFT.

In this paper, we have proposed a conceptually and computationally simple method to treat memory effects
in TDDFT. Our approach is based on a new extension of the well-known
generator coordinate (or Griffin-Hill-Wheeler) method to time-dependent systems. In essence, we represent
the $N$-body wavefunction as a superposition of a few nonorthogonal KS Slater determinants, coming from a set of
TDKS Hamiltonians featuring deformed XC potentials. The weight functions are optimized with a stationary
action principle. The key is that our method allows not only {\em spatial}, but also {\em temporal}
deformations of the XC potential. This allows for a very cheap way to bring in memory,
namely by evaluating the XC potential at time $t$ using a density at a previous time $t_r$ as input.
The variational principle then determines how each of these retarded XC potentials are weighted relative
to each other. The numerical effort of the resulting GC-TDDFT scheme is quite manageable.

We have applied our GC-TDDFT scheme to describe parametric oscillations in a strongly driven Hooke's atom,
a two-electron system with the special properties that it cannot ionize and that its levels are nearly equally
spaced. Therefore, the interplay between single and multiple excitations is particularly pronounced, leading
to a very strong modulation of the free charge-density oscillations. These modulations are impossible to
capture in ALDA or any non-retarded GC-TDDFT scheme. By contrast, our GC-TDDFT with memory reproduced the
effect extremely well.

The GC-TDDFT allows for a wide range of possible deformations in the TDKS Hamiltonian. As we have shown,
one can choose deformations which remove part of the self-interaction error, and which bring in retardation
in a simple way. The success of the GC-TDDFT approach thus crucially depends on choosing deformations according to the
particular physical characteristics and requirements of the systems under study. We believe that, due to
its flexibility and computational simplicity, the GC-TDDFT approach may prove valuable for a wide range of
applications in chemistry and physics.

\section{Acknowledgements}
This work was supported by FAPESP, CAPES and CNPq.
E. O. wishes to thank the University of Missouri-Columbia for its hospitality.
C. A. U. is supported by NSF Grant No. DMR-0553485 and by Research Corporation.

\appendix

\section{Properties of the TDGHW equation} \label{AppendixA}
The static GHW equation (\ref{GHWeq}) follows from the variational minimization of
$E=\langle \Psi | \hat{H} | \Psi \rangle/\langle \Psi| \Psi \rangle$.
With the Hamiltonian and overlap matrix elements of the given DFT seed functions,
$K(\alpha,\alpha')=\langle\Phi_{\rm KS}(\alpha)|\hat{H}|\Phi_{\rm KS}(\alpha')\rangle$ and
$S(\alpha,\alpha')=\langle\Phi_{\rm KS}(\alpha)|\Phi_{\rm KS}(\alpha')\rangle$, we can write this as
\begin{equation}
\frac{\delta}{\delta f^*(\alpha)}
\frac{\int d\beta \int d\gamma f^*(\beta) K(\beta,\gamma) f(\gamma)}{\int d\beta \int d\gamma f^*(\beta) S(\beta,\gamma)
f(\gamma)} = 0 \:.
\end{equation}
The resulting static GHW equation is given by
\begin{equation}\label{GHWappendix}
\int d\alpha'\left[ K(\alpha,\alpha') - ES(\alpha,\alpha') \right] f(\alpha')=0 \:.
\end{equation}
The static GHW equation provides the initial condition for the subsequent time propagation with the TDGHW scheme.
From the stationary-action principle
\begin{eqnarray}
\hspace*{-1cm}
0 &=&
\frac{\delta}{\delta f^*(\alpha,t)} \int_{t_0}^{t_1} \!dt' \!
\int \!d\beta \!\int d\gamma \: f^*(\beta,t')\nonumber\\
&\times& \!\! \left\langle \Phi_{\rm KS}(\beta,t')\left| i\frac{\partial}{\partial t'} - \hat{H}(t')\right|
\Phi_{\rm KS}(\gamma,t')\right\rangle \! f(\gamma,t')
\end{eqnarray}
we obtain the TDGHW equation
\begin{equation} \label{TDGHWappendix}
\int d\alpha'\left[ A(\alpha,\alpha',t) + S(\alpha,\alpha',t)i\frac{\partial}{\partial t} \right]
f(\alpha',t)=0,
\end{equation}
where the action and overlap matrix elements $A$  and $S$ are defined in eqs. (\ref{actionmatrix}) and (\ref{overlapmatrix}).

\subsection{Static limit}
Let us first show that the TDGHW equation reduces to the static GHW equation in the special case where the external
one-body potential is time-independent. In that case, the KS Slater determinant associated with the
deformation parameter $\alpha$ becomes
\begin{equation}
\Phi_{\rm KS} (\alpha,t) \longrightarrow e^{-iE^\alpha_{\rm KS}t} \Phi_{\rm KS}(\alpha)
\end{equation}
where $E^\alpha_{\rm KS} = \sum_{j=1}^N \varepsilon^\alpha_i$ is the total KS energy of the system.
Equation (\ref{TDGHWappendix}) then becomes
\begin{eqnarray}
0 &=&
\int d\alpha' e^{-iE^{\alpha'}_{\rm KS}t}\bigg[ E_{\rm KS}^{\alpha'}S(\alpha,\alpha') -
K(\alpha,\alpha') \nonumber\\
&& {}+ S(\alpha,\alpha')i\frac{\partial}{\partial t} \bigg]
f(\alpha',t)\:,
\end{eqnarray}
which reduces to the static GHW equation (\ref{GHWappendix}) if
\begin{equation}
f(\alpha,t) \longrightarrow e^{-i(E-E_{\rm KS}^\alpha)t} f(\alpha) \:.
\end{equation}
Thus, for a time-independent Hamiltonian $\hat{H}$ the weight functions only have a simple time-dependent
phase factor, and we obtain $\Psi(t) \longrightarrow e^{-iEt} \Psi$ as expected.

\subsection{Norm conservation}
The derivation of the TDGHW equation (\ref{TDGHWappendix}) implicitly
assumes that the initial many-body wave functions are normalized, and that the norm
\begin{equation}
\langle \Psi(t) | \Psi(t) \rangle = \int \! d\alpha \! \int\!  d\alpha' f^*(\alpha,t) S(\alpha,\alpha',t)f(\alpha',t)
\end{equation}
remains constant for all times. Indeed, using the definition of $S(\alpha,\alpha',t)$ and eq.
(\ref{TDGHWappendix}), it is straightforward to show that
\begin{equation}
\frac{d}{dt} \langle \Psi(t) | \Psi(t) \rangle = 0 \:,
\end{equation}
which establishes norm conservation of the TDGHW scheme.

\section{Numerical solution of the TDGWH equation} \label{AppendixB}
By discretizing the generator coordinate $\alpha$, the GHW and TDGHW integral equations are turned
into linear algebra problems. The static GHW equation becomes a generalized eigenvalue problem \cite{Capelle2003}:
\begin{equation}\label{GHWdiscrete}
\sum_{\alpha'}\left[ K_{\alpha,\alpha'} - ES_{\alpha,\alpha'} \right] f_{\alpha'}=0 \:.
\end{equation}
Next, we consider the TDGHW equation, where from now on we omit the time argument for notational simplicity:
\begin{equation}\label{TDGHWdiscrete}
\sum_{\alpha'}\left[ A_{\alpha,\alpha'} + S_{\alpha,\alpha'} \:i\frac{\partial}{\partial_t} \right]
f_{\alpha'}=0
\end{equation}
which, as we have shown above, conserves the norm
\begin{equation}\label{norm1}
N = \sum_{\alpha \alpha'} f^*_\alpha S_{\alpha\alpha'} f_{\alpha'} \:.
\end{equation}
Calculation of the square root of the overlap matrix $\bf S$, defined as
\begin{equation}
S_{\alpha \alpha'} = \sum_\gamma R_{\alpha \gamma} R_{\gamma \alpha'} \:,
\end{equation}
is a standard task in linear algebra. Both $\bf S$ and $\bf R$ are hermitian matrices.
Thus, the norm (\ref{norm1}) can be rewritten as
\begin{equation}
N = \sum_\gamma F^*_\gamma F_\gamma \:,
\end{equation}
where
\begin{equation}
\vec{F} = {\bf R} \vec{f} \:.
\end{equation}
The advantage of working with the vector $\vec{F}$ is that it satisfies a formally simpler normalization
condition compared to the original vector $\vec{f}$ of the weight functions, which involves the overlap matrix.
It is straightforward to transform the TDGHW equation (\ref{TDGHWdiscrete}) into an equation of motion
for $\vec{F}$:
\begin{equation} \label{Feq}
\left[ {\bf M}+ i\frac{\partial}{\partial t}\right] \vec{F} = 0
\end{equation}
where
\begin{equation}
{\bf M} = {\bf R}^{-1} {\bf A} {\bf R}^{-1} - \left(i\frac{\partial }{\partial t} {\bf R}\right) {\bf R}^{-1} \:.
\end{equation}
Using ${\bf R}^\dagger = {\bf R}$ and ${\bf A}^\dagger = {\bf A}- i\partial {\bf S}/\partial t$ it
can be shown that ${\bf M}^\dagger = {\bf M}$.

The numerical solution of the equation of motion (\ref{Feq}) for $\vec{F}$ can now be carried out using
the standard Crank-Nicholson algorithm \cite{Nrecipes}, which is unitary and accurate to second order in the time step:
\begin{equation}
\left[ {\bf 1} + \frac{i\Delta t}{2} {\bf M}\right] \vec{F}(t+\Delta t)
= \left[ {\bf 1} - \frac{i\Delta t}{2} {\bf M}\right] \vec{F}(t) \:,
\end{equation}
where $\bf M$ is evaluated at time $t + \Delta t/2$. Once the time evolution of $\vec{F}$ has been calculated,
it is a simple matter to obtain the time-dependent weights $\vec{f}={\bf R}^{-1}\vec{F}$.

\end{document}